\documentclass{article} 
\usepackage{epsfig,amssymb,euscript} \usepackage{amsmath}

\newcommand{\ket}{\rangle} 

\newcommand{\be}{\begin{equation}} \newcommand{\ee}{\end{equation}}
\newcommand{\bea}{\begin{eqnarray}} \newcommand{\eea}{\end{eqnarray}}
\newcommand{\ba}{\begin{array}} \newcommand{\ea}{\end{array}}

  \def\bbox{{\,\lower0.9pt\vbox{\hrule \hbox{\vrule height 0.2 cm
\hskip 0.2 cm \vrule height 0.2 cm}\hrule}\,}} \newcommand{\dsl}{\pa
\kern-0.5em /}

\def\ds{\raise.15ex\hbox{/}\kern-.57em\partial}
\def\Ds{\,\raise.15ex\hbox{/}\mkern-13.5mu D}

\begin{document}

\makeatletter
\renewcommand{\theequation}{\thesection.\arabic{equation}}
\@addtoreset{equation}{section} \makeatother

\baselineskip 18pt


\begin{titlepage}

\vfill

\begin{center}
\baselineskip=16pt {\Large\bf Simulating Causal Wave Function Collapse Models}
\vskip 10.mm Fay Dowker$^{a}$ and Isabelle Herbauts$^{b}$\\
\end{center}  
\vfill \par
\begin{center}
{\bf Abstract}
\end{center}
\begin{quote}
We present simulations of causal dynamical wave function collapse
models of field
theories on a $1+1$ null lattice.  We use our simulations to compare
and contrast two possible interpretations of the models, one in which
the field values are real and the other in which the state vector is
real. We suggest that a procedure of coarse graining and renormalising
the fundamental field can overcome its noisiness and argue that
this coarse grained renormalised field will show interesting structure
if the state vector does on the coarse grained scale. We speculate on the
implications for quantum gravity. 

\end{quote}

\vfill

\hrule width 5.cm \vskip 5mm {\small
\noindent $^a$ Blackett Laboratory, Imperial College, London SW7 2BZ,
UK and Perimeter Institute, Waterloo, Ontario N2T 2A9, Canada.\\
\noindent $^b$ Department of Physics, Queen Mary, University of
London, London E1 4NS, UK.\\ }
\end{titlepage}
\setcounter{equation}{0}

\section{Introduction}

The tension between relativity and quantum theory exists even before
one tries to include gravity in the picture and even if one tries to
take a determinedly operationalist stance one is forced to try to make a
careful statement of what quantities may be measured and how
({\it e.g.} \cite{Sorkin:1993gg, Beckman:2001ck}).  
If one is seeking an observer
independent alternative to standard quantum theory, however,
then that tension escalates into an issue of fundamental importance,
whose resolution may lead to the ``radical conceptual
renewal'' anticipated by John Bell \cite{Bell:1987i}.  This conceptual
advance may be intimately linked with progress in finding a quantum
theory of gravity, for example by finding the correct, quantum
analogue of the ``Bell Causality'' condition for causal sets
\cite{Rideout:1999ub} but we have a more modest objective in this
paper. We will investigate a simple observer independent dynamical
collapse model on a lattice which is causal in the sense that external
agents cannot manipulate it to communicate faster than the lattice
causal structure ought to allow \cite{Dowker:2002wm}.

The model in question is a model for a discrete quantum field theory
and it falls in the general category of ``dynamical localisation''
models (for a substantial review see \cite{Bassi:2003gd}). It can be
thought of as a causal, discrete Continuous Spontaneous Localisation
(CSL) model \cite{Diosi:1988a, Pearle:1989, Ghirardi:1990}.  
Dynamical collapse
models admit (at least) two possible interpretations.  In the first,
which we call the State Interpretation (SI) and which is used by
almost all workers in this area, it is the state vector which is real
(see for example \cite{Pearle:1998yf}). In the second, which we call
the Field Interpretation (FI), it is the field values which are real
(in the original GRW model \cite{Ghirardi:1986mt} the analogous
interpretation first proposed by Bell \cite{Bell:1987ii} and
championed by Kent \cite{Kent:1989nk, Kent:1998bc} is that the
collapse centres are real). 

This choice of interpretation is not something that is often
highlighted. We believe that it is a physical choice and it is one of
the purposes of the current paper to investigate the consequences of
the choice.  It may be that SI and FI are operationally
indistinguishable in which case there is less motivation for
considering one over the other, though each may suggest different
directions for future research. It may be that the interpretations
lead to verifiably different predictions which would be an extremely
interesting result. So far, almost all work on collapse models and
comparisons of their implications with experiment have assumed the
SI. We would like to take a small step towards determining whether the
FI has anything interesting to say about the real world by
investigating it in the context of our lattice model.

The paper is structured as follows. In section 2 we review the lattice
collapse models of \cite{Dowker:2002wm}.  In section 3 we describe our
simulations of the SI and the FI, for a range of parameters for some
simple initial states. In section 4 we investigate the decay of
superpositions. Section 5 is a description of coarse graining and
renormalisation in the model. Section 6 contains a discussion of
parameters, a description of future work and speculations on
applications to quantum gravity.

\section{1+1 lattice collapse models}

In order to motivate the lattice collapse models, we first review the 
GRW model, following the presentation of Kent \cite{Kent:1998bc} 
who calls it the ``ur-model'' of modern dynamical 
collapse models. In the  GRW model 
for a single spinless particle in one dimension the wave function
$\psi( x , t )$
undergoes two types of evolution.  Almost all of the time, it follows
the Schr\"odinger equation, but at discrete randomly chosen times 
it jumps discontinuously, so that 
\be
\psi \rightarrow \frac{ J_{\hat{x}} \psi } 
{\left(\int dx |J_{\hat{x}}\psi |^2\right)^{\frac{1}{2}}},  
\ee
where $J_{\hat{x}}$ is the function of $x$
\be
J_{\hat{x}}(x) = a^{-\frac{1}{2}}\pi^{-\frac{1}{4}} e^{-(x-\hat{x})^2/2a^2}\;.
\ee
The 
 $\hat{x}$ is chosen randomly according to the probability 
distribution
\be
{\text Prob}(\hat{x})= \int d x  |J_{\hat{x}} \psi |^2 
\, 
\ee
which is properly normalised to be a probability distribution 
because
\be
\int d \hat{x} J_{\hat{x}} (x)^2 = 1 \;.
\ee
$a$ is a constant parametrising the model.   
The times of the jumps are given by a Poisson process, with
mean interval $\tau / N$ between jumps.  The parameters $\tau$ and $a$
are to be thought of as new constants of nature; GRW originally
suggested
\be
a \approx 10^{-5} {\text~cm} , \qquad 
 \tau \approx 10^{15} {\text~sec}.
\ee
In the interpretation of the GRW model due to Bell, it is the sequence of 
stochastically generated events -- $(\hat{x},t)$'s --  
which constitutes reality in the model and the model
assigns a probability distribution to the sample space of all such sequences,
$(\hat{x}_1, t_1), \dots (\hat{x}_n, t_n)$, which is the probability of 
that sequence of times (Poisson distributed) multiplied by
\be
\Vert J_{\hat{x}_n} U(t_n, t_{n-1}) \dots J_{\hat{x}_1} U(t_1, t_0) \psi(t_0)
\Vert^2
\ee
where $U(t_{i+1}, t_i)$ is the standard Schr\"odinger time evolution operator
between two collapse times. 

The lattice collapse models aim to emulate the formalism of the GRW 
model in the context of a field theory with non-trivial causal structure. 
They are based on light cone lattice field theories in 1+1
dimensions, introduced in the study of integrable models
\cite{Destri:1987ze}.  We follow the presentation of Samols
\cite{Samols:1995zz} of this ``bare bones'' local quantum field
theory.  Spacetime is a $1+1$ null lattice, periodically identified in
space with $N$ spatial vertex sites.  The links of the lattice
are left or right going null rays.

\begin{figure}[thb]
\centerline{\epsfbox{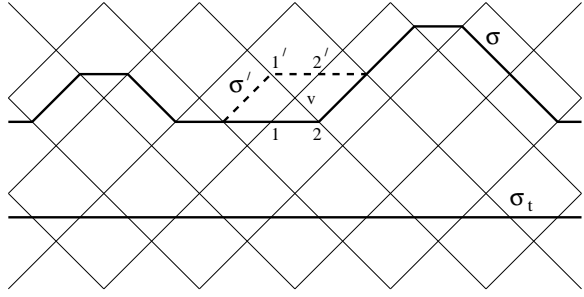}}
\caption{ The light-cone lattice. $\sigma_t$ is a constant time slice;
$\sigma$ is a general spacelike surface and $\sigma'$ one obtained
from it by an elementary motion across the vertex $v$. Links labelled
$1$ and $2$ ($1'$ and $2'$) are the ingoing (outgoing) links at $v$.}
\label{lattice:fig}
\end{figure}

A spacelike surface, $\sigma$, is given by the set of links cut by the
surface and is specified completely by the position of an initial link
and a sequence of $N$ right-going and $N$ left-going 
links, moving from left to right, starting with the initial
link. An example is shown in figure \ref{lattice:fig}.

The local field variables, $\alpha$, live on the links. At link $l$
the variable $\alpha_l$ takes just two values, $0$ or $1$, and there
is a (``qubit'') Hilbert space, ${\cal{H}}_l$, spanned by two states
labelled by $\alpha_l=0$ and $\alpha_l=1$.  At each vertex, $v$, the
local evolution law is given by a 4-dimensional unitary R-matrix,
$U(v)$, that evolves from the 4-d Hilbert space that is the tensor
product of the Hilbert spaces on the two ingoing links to the 4-d
Hilbert space on the two outgoing links.

A quantum state $|\Psi_\sigma\rangle$ on a spacelike surface,
$\sigma$, is an element of the $2^{2N}$-dimensional Hilbert space,
${\cal{H}}_\sigma$, that is the tensor product of the Hilbert spaces
on all the links cut by $\sigma$. The $\alpha$ basis vectors of
${\cal{H}}_\sigma$ (the only ones we will use: they form the preferred
or physical basis) are labelled by the possible field configurations
on $\sigma$, namely the $2N$-element bit strings $\{0,1\}^{2N}$.  We
identify the Hilbert spaces on different surfaces in the
obvious way by identifying 2-dimensional Hilbert spaces on links
vertically above and below each other on the lattice.

In the standard quantum theory, 
the unitary evolution of the state to another spacelike surface
$\sigma'$ is effected by applying all the R-matrices at the vertices
between $\sigma$ and $\sigma'$, in an order respecting the causal
order of the vertices.  In the simplest case, when only a single
vertex is crossed (to the future of $\sigma$) the deformation of
$\sigma$ to $\sigma'$ is called an ``elementary motion'' and an
example is shown in figure \ref{lattice:fig}.  The R-matrices are left
unspecified for now to keep the discussion as general as possible.  In
a conventional field theory, they will be uniform over the lattice.

We assume that there is an initial spacelike surface, $\sigma_0$, with
a state, $|\Psi_0\rangle$, on that surface. We consider the lattice to
extend into the infinite future. The lattice vertices to the future of
$\sigma_0$ are partially ordered by their spacetime causal order. We
define a {\it stem} to be a finite subset of the vertices to the
future of $\sigma_0$ which contains its own past (to the future of
$\sigma_0$).  Consider a {\it natural labelling} of all the vertices
to the future of $\sigma_0$: $v_1, v_2, v_3 \dots$ so that if $v_i$ is
in the causal past of $v_j$ then $i < j$. This is a {\it linear
extension} of the partial order and the finite set $\{v_1, v_2 \dots
v_k\}$ is a stem for any $k$. We also introduce the notation
$\sigma_k$ to denote the spacelike surface reached after the
elementary motions across $v_1, v_2 \dots v_k$ have been completed. There
is a one-to-one correspondence between stems and spacelike surfaces to
the future of $\sigma_0$.

On each link ({\i.e.} on each 2-dimensional Hilbert space associated
with a link) we consider the two jump operators $J_0$ and $J_1$ where
\be J_0 = \frac{1}{\sqrt{1+ X^2}}\begin{pmatrix} 1&0\\ 0&X\\
\end{pmatrix}\;, \quad J_1 = \frac{1}{\sqrt{1+ X^2}}\begin{pmatrix}
X&0\\ 0&1\\
\end{pmatrix}\;,
\label{eq:jumps}
\ee
and $0\le X\le 1$. 
These would be projectors onto states $|0\rangle$ and $|1\rangle$
respectively if $X=0$. They satisfy $J_0^2 + J_1^2 = 1$ and are known
as ``Kraus'' operators.

It is convenient to introduce the notation $\hat{\alpha}_{v_k}$ to
denote the values of the $\alpha$ field variables on the two outgoing
links (one left and one right) from the vertex $v_k$.  Then the
possible values of $\hat{\alpha}_{v_k}$ are $\{(0,0), (0,1), (1,0),
(1,1)\}$ and we define the jump operator $J(\hat{\alpha}_{v_k})$ on
the 4-dimensional Hilbert space on the outgoing links from $v_k$ as
the tensor product of the two relevant 2-dimensional jump operators,
{\it e.g.} when $\hat{\alpha}_{v_k} = (0,0)$, $J(\hat{\alpha}_{v_k})=
J_0 \otimes J_0$. We promote $J(\hat{\alpha}_{v_k})$ to an operator on
the Hilbert space of any spatial surface containing those two links by
taking the tensor product with the identity operators on all the other
components of the full Hilbert space and, finally, we define a
``Heisenberg picture'' operator
\be J_{v_k}(\hat{\alpha}_{v_k}) \equiv U^{-1}(v_1)\dots U^{-1}(v_k)
J(\hat{\alpha}_{v_k}) U(v_k)\dots U(v_1)\; .  \ee
The probability of the field configuration $\{\hat{\alpha}_{v_1},
\dots \hat{\alpha}_{v_n}\}$ is given by
\be\label{probability.eq} P(\hat{\alpha}_{v_1}, \dots
\hat{\alpha}_{v_n}) = \Vert J_{v_n}(\hat{\alpha}_{v_n})\dots
J_{v_1}(\hat{\alpha}_{v_1}) |\Psi_0 \rangle \Vert^2\; .  \ee
This depends only on the (partial) causal order of the vertices
because any other choice of natural labelling of the same vertices
gives the same result (because the operators at spacelike vertices
commute). It is only the limitation of our mathematical notation that
means we have to specify a linear extension to write down the formula
\eqref{probability.eq}.  These probabilities of the field
configurations on all the stems are enough, via the standard methods
of measure theory, to define a unique probability measure on the
sample space of all field configurations on the semi-infinite lattice.

The probability \eqref{probability.eq} can also be written in terms of the
``Schr\"odinger Picture'' operators $J({\hat{\alpha}}_{v_k})$
\be\label{probschr.eq} P(\hat{\alpha}_{v_1}, \dots
\hat{\alpha}_{v_n}) = \Vert J(\hat{\alpha}_{v_n})U(v_n)\dots
  J(\hat{\alpha}_{v_1}) U(v_1)|\Psi_0\rangle  \Vert^2 \;.
\ee
The {\it state} on the surface $\sigma_n$ that is reached after
the elementary motions over vertices $v_1, \dots v_n$ and the field
values $\{\hat{\alpha}_{v_1}, \dots \hat{\alpha}_{v_n}\}$ have been
realised is the normalised state
\be\label{state.eq} |\Psi_n\rangle =
\frac{J(\hat{\alpha}_{v_n})\dots J(\hat{\alpha}_{v_1})
|\Psi_0\rangle } {\Vert J(\hat{\alpha}_{v_n})\dots
J(\hat{\alpha}_{v_1}) |\Psi_0\rangle \Vert} \;  \ee
where this is given in terms of the Schr\"odinger operators, 
$J(\hat{\alpha}_{v_k})$. Note that, given a particular 
realised field configuration on the whole lattice, the state on
any arbitrary spacelike surface can be calculated and it is 
independent of the linear ordering of the vertices to its
past that is chosen in order to write it down.

The Field Interpretation FI is captured by the slogan, ``the field
values are real''. Reality is rooted in spacetime. This is a covariant 
interpretation because nothing depends on the linear order 
chosen to express the probability: the joint probability distribution on the
field values on the lattice depends only on the causal order of the
vertices.  Any total ordering used in the analysis is entirely
fictitious and unphysical. One can maintain a ``block universe''
perspective in which the reality is the whole semi-infinite lattice
to the future of $\sigma_0$ with a field configuration on it. 
It is also possible, however, to have a more ``dynamic'' picture of
the model in which the ``events'' (realisations of field
values on the outgoing links of a vertex) {\it occur} but
with the order in
which they occur physically being a {\it partial} order.

The essence of the State Interpretation SI is, ``the state is
real''. The fundamental arena of reality is Hilbert space. 
Given a particular realisation of the field on the entire 
lattice, there is a unique state associated with each spacelike 
surface. The SI is covariant in the sense 
that the state on a spacelike surface 
is independent of the linear order of the vertices to its past 
chosen to write down the state mathematically (as in \eqref{state.eq}). 
However, as we will see, defining {\it local}
quantities on the lattice -- which is necessary in order to 
make predictions -- will bring in a dependence on hypersurfaces.

We have described two possible interpretations of the formalism, SI
and FI. There is another, suggested by Lajos Di\'osi and
Gerard Milburn \cite{priv:2003}, which is that it describes, in the
standard (no-collapse) quantum theory, an open system coupled to local
ancilla systems on each link which are themselves measured.  The
field values then label the particular outcomes of these ancillae
measurements. Alternatively, if the ancilla states are traced over,
this would give rise to a density matrix, identical to that obtained
by summing $|\Psi\rangle\langle\Psi|$ (where $|\Psi\rangle$ is given
by \eqref{state.eq}) over the field configurations weighted by their
probabilities. The individual histories we consider in FI and SI would
then be thought of as a particular ``unravelling'' of the evolution of
this reduced density matrix. Since the lattice field theories (without
collapse) described here have been interpreted as ``quantum lattice
gas automata'' (QLGA) \cite{Meyer:1996kt} we could therefore interpret
the present work as investigations into the effects of environmental
decoherence on these interesting proposed architectures for quantum
computation.
 
\section{Comparing the SI and FI for various parameters}

The field variables on the lattice links
take only two values, $\alpha_l=\{0,1\}$. These values can be thought
of as occupation numbers, so that an
``occupied'' $|1\ket$ or ``empty'' $|0\ket$ state can be associated to
each link. The diagonal links of the lattice are the possible
world-lines for propagation forwards in time of ``bare'' particles,
moving either right or left and the lattice consists therefore of
an array of nodes that can be occupied by left or right moving
particles.  These bare particles would not however, become the
physical particles in a eventual continuum theory
\cite{Destri:1987ze}.

The local evolution law at each vertex on the lattice is encoded in a
4-dimensional unitary $R$-matrix, whose entries are the amplitudes
connecting the four possible states on the ingoing and outgoing pairs
of links at that vertex. Different conservation laws can be imposed on
the dynamics of a system by a suitable choice of the $R$-matrix,
describing the different processes that can take place on the lattice.

For a field theory with spacetime translation invariance, the
$R$-matrices are also chosen to be uniform over the lattice.  One
special choice is a particle number preserving matrix, i.e. a
$R$-matrix conserving occupation numbers at each vertex. Such a matrix
can be parametrised as:
\be{ U = \bordermatrix{ &
 \scriptscriptstyle{\phantom{\nearrow\nwarrow}} &
 \scriptscriptstyle{\phantom{\nearrow}\nwarrow} &
 \scriptscriptstyle{\nearrow\phantom{\nwarrow}} &
 \scriptscriptstyle{\nearrow\nwarrow} \cr
 \scriptscriptstyle{\phantom{\nwarrow\nearrow}} & 1 & 0 & 0 & 0 \cr
 \scriptscriptstyle{\phantom{\nwarrow}\nearrow} & 0 & i e^{i
 \alpha}\sin\theta & e^{i \alpha}\cos\theta & 0 \cr
 \scriptscriptstyle{\nwarrow\phantom{\nearrow}} & 0 & e^{i
 \alpha}\cos\theta & i e^{i \alpha}\sin\theta & 0 \cr
 \scriptscriptstyle{\nwarrow\nearrow} & 0 & 0 & 0 & e^{i \beta} \cr }
 \; .
\label{eq:rmat}
} \ee
 The local unitary evolution given by the $R$-matrix \eqref{eq:rmat}
 imposes both particle number conservation and parity invariance
 (symmetry under left-right exchange) throughout the lattice. An
 $R$-matrix of this type leads to a particular fermionic model, the
 massive Thirring model, in a suitable continuum limit of the unitary
 (no collapse) theory \cite{Destri:1987ze}.

Each of the six non-zero amplitudes can be thought of as indicating
possible paths for a particle at each vertex. These paths are
symbolically indicated in equation \eqref{eq:rmat} by $\nwarrow$ for a
left moving particle, $\nearrow$ for a right moving particle.  Without
loss of generality, the ``nothing to nothing'' amplitude is chosen to
be 1.  The ``interaction'' term between two ingoing particles is a
phase multiplication only, $e^{i \beta}$.  For the simulations
presented in this paper, we set $\alpha$ and $\beta$ to $0$.

Variation of $\theta$ controls the average speed of the particles on the
lattice: if $\theta=\pi/2$ the R-matrix is proportional to the
identity matrix, and the particles change direction at each vertex and
there is no propagation in space and the particles have infinite
``mass''. At the other extreme, with $\theta =0$ the R-matrix interchanges
the states on the two incoming links, so that the particles never
change direction and follow null lines on the lattice and can be
thought of as massless. 

In our simulations we are limited as to lattice size by the
exponential growth of the problem in vertex number. For a lattice with
$N$ vertices, this means handling vectors with $2^{2N}$ components,
and dealing with $N$ multiplications of these vectors by matrices of
dimensions $2^{2N} \times 2^{2N}$ at each time step (each constant
time slice). We ran simulations for $N = 8$, 9 and 10.

The simulation proceeds in stages as follows. Given the evolution up to 
surface $\sigma_n$ (in other words we 
have the surface, realised field values at all vertices 
between $\sigma_0$ and $\sigma_n$ and the normalised state on $\sigma_n$,
$|\Psi_n\rangle$, depending on those past realised values
according to \eqref{state.eq}), 
an elementary motion is chosen at random from 
those possible with equal probability. The motion is implemented
resulting in a new hypersurface $\sigma_{n+1}$ and
and the state is unitarily evolved to a state on $\sigma_{n+1}$
\be
|\tilde{\Psi}_{n+1}\rangle = U(v_{n+1}) |\Psi_n\rangle \;.
\ee
Field values at the new vertex $v_{n+1}$ (strictly, on the two 
outgoing links from the new vertex) are chosen at random 
according to the probability distribution   
\be
{\text Prob}(\hat{\alpha}_{n+1}) = \Vert J(\hat{\alpha}_{n+1}) 
|\tilde{\Psi}_{n+1}\rangle \Vert^2\;.
\ee
This is the conditional probability distribution on field values
at $v_{n+1}$ given the field values that have already been 
realised to the past of $\sigma_n$. 

If the field value $\hat{\alpha}_{n+1}$ is realised then 
the state is ``hit'' and becomes the new state on $\sigma_{n+1}$
\be 
|\Psi_{n+1} \rangle = \frac{ J(\hat{\alpha}_{n+1}) 
|\tilde{\Psi}_{n+1}\rangle}{\Vert J(\hat{\alpha}_{n+1}) 
|\tilde{\Psi}_{n+1}\rangle \Vert^2}
\ee
and the process repeats.

Note that this evolution scheme is tantamount to putting a 
specific probability distribution on linear orderings of the vertices --
fixed by the rule for choosing elementary motions
at random with equal probability from those possible at 
each stage -- and choosing
one particular linear ordering according to that probability 
distribution. The distribution we have chosen
is generated by a Markovian rule for 
evolution and thus is particularly easy to simulate. Other rules can 
be chosen -- for example, one might choose to weight all linear 
orderings equally. Due to the fact that the probability distribution
on the field configurations depends only on the causal order of the 
vertices, the results of the simulation for the field values 
do not depend on the choice of distribution for the linear order: {\it any}   
choice will lead to a correct simulation of the physical distribution 
on field configurations (we could even choose a fixed linear ordering 
with, for example, increasing labels from left to right, row by row).
In the simulation of the evolution of the {\it state}, however, the 
probability distribution on linear orderings does play a role because
a linear ordering is equivalent to a sequence of surfaces and in the
simulation, the state is only calculated on the surfaces that are 
included in the chosen sequence. Even so, the state can still be 
thought of as covariant because it is determined on every surface by the
realised field configuration, 
even though our simulation doesn't actually calculate it. 

\begin{figure}[thb]
\epsfxsize=6cm \centerline{\epsfbox{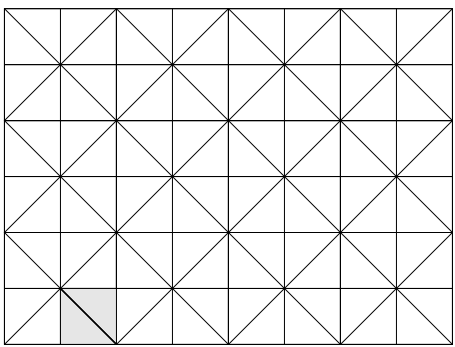}}
\caption{ The cells on the light-cone lattice. The lattice is
represented by an array of square cells, each cell containing one link
on a diagonal.  One such cell is shown by the shaded area.}
\label{cell:fig}
\end{figure}

We will show two types of plots. In both of them each cell corresponds to a
single link of the lattice (so that there are two cells per vertex),
as shown in figure \ref{cell:fig}.

In one type of plot (FI) the cell is white if the field value is 0 and
black if the field value is 1. In the second type (SI) the darkness of
the cell is (positively) proportional to the square of the amplitude
(according to the state at that stage of the evolution) 
for the field value 1 on the link. (in other
words the plot is of ``stuff'' \cite{Pearle:1998yf}).  It is important
to emphasize that this SI plot is {\it not} the probability that the
field value on the link will be 1. Indeed, let the link in question 
be $l$ and suppose, at
some stage in the dynamics, $l$ is one of the outgoing
links from the vertex that has just been evolved over. Let the state
on the current spacelike surface through $l$ be denoted schematically
by
\be \label{current.eq}
\vert \Psi \rangle = a \vert 0 \rangle + b \vert 1 \rangle \ee
where $\vert 0 \rangle$ ($\vert 1 \rangle$) is short hand for the
normalised superposition of all the terms in the state in which the
value of the field on $l$ is 0 (1).  The probability that the field
will be 1 on $l$ (conditional on the past evolution to that stage) is
\be \label{probone.eq} \frac{|a|^2 X^2 + |b|^2}{1 + X^2}\; ,   \ee
whereas the square of the amplitude for field value 1, which 
we show in the SI plot, is $|b|^2$.

Here, in defining this {\it local} quantity on the lattice that we can 
plot and analyse,  we have introduced a dependence on 
spacelike surfaces (and hence on the choice we have 
made of a probability distribution on sequences of 
surfaces). Since, although the state on a surface is a covariant 
quantity, the value of $|b|^2$ on a {\it link} is 
not covariant because it depends on the state on an entire surface 
through that link: choose a different surface and the value
of $|b|^2$ will be different (due to the non-unitarity of the hits). 
In our simulation, the surface through the link is whatever it 
happens to be at that stage of the evolution -- in other words the 
surface is chosen at random according to the probability distribution 
on surfaces defined by the Markovian evolution rule for elementary 
motions. So our state plot of the quantity $|b|^2$ depends on that 
choice. A different choice ({\it e.g.} equally weighted linear 
orderings of vertices, or one fixed deterministic ordering) would 
give a different state plot. Alternatively, a covariant choice for the
surface to use in calculating $|b|^2$ on a link
-- not equivalent to any kind of ``evolution'' 
law for surfaces, stochastic or otherwise --  would be the boundary of the 
causal past of the vertex from which the link is outgoing. 
We have not investigated these alternatives.

\begin{figure}[thb]
\epsfxsize=16cm \centerline{\epsfbox{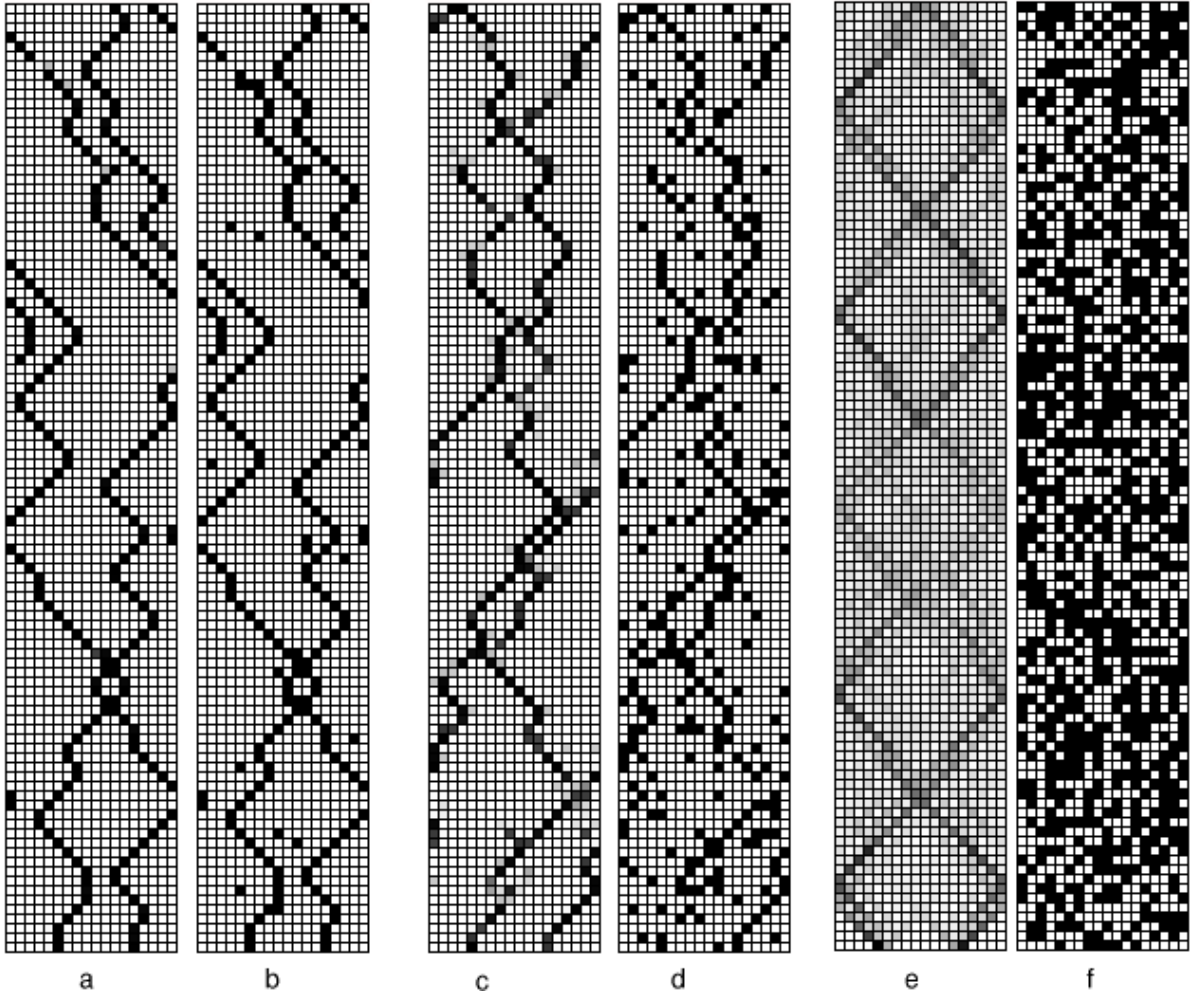}}
\caption{Pairs (a,b), (c,d) and (e,f), of (SI,
FI) plots for three different runs of the 
simulation.  The initial state is a field eigenstate
with two particles and $\theta=\pi/6$ in all three pairs. The pairs
have different values of $X$: $X=0.1$ for (a,b), $0.3$ for (c,d) and
$0.95$ for (e,f).}
\label{fig:2part}
\end{figure}

In figure \ref{fig:2part} the plots are grouped in pairs (a,b), (c,d)
and (e,f) of (SI, FI) plots.
The members of a pair have the same parameters and the plots in
the pair refer to exactly the same run.  The initial state
is a field eigenstate with two particles and $\theta=\pi/6$ in all
three pairs. The pairs have different values of $X$: $X=0.1$ for
(a,b), $0.3$ for (c,d) and $0.95$ for (e,f).
The plots all begin on the row of links immediately above the initial surface 
because there are no field values on
the initial surface itself.

We can see that when the jump operators are close to being projectors
the state remains very close to being in an eigenstate: the R-matrices
introduce superpositions which are essentially killed at each step by
the jump operators. The two plots (a) and (b) in 
figure \ref{fig:2part} are virtually identical, as they
should be and the trajectories seen are very close to being 
classical random walks. Indeed, if $X$ were exactly 0, they would
be classical (non-Markovian) random walks where, at each vertex,
 the walker either continues in the null 
direction it is going in with probability 
$\cos^2 (\pi/6) = 3/4$ or changes direction with
probability $\sin^2 (\pi/6) = 1/4$.

 As $X$ is increased with fixed $\theta$, the balance changes and 
more signs of superposition appear in the
SI plot whilst the field plot starts to look much more noisy. For $X =
0.95$, the state plot resembles the unitary dynamics shown in
the right hand plot of figure 9 in \cite{Meyer:1996kt} and the field
plot has no discernable structure any more.

\begin{figure}[thb]
\epsfxsize=16cm \centerline{\epsfbox{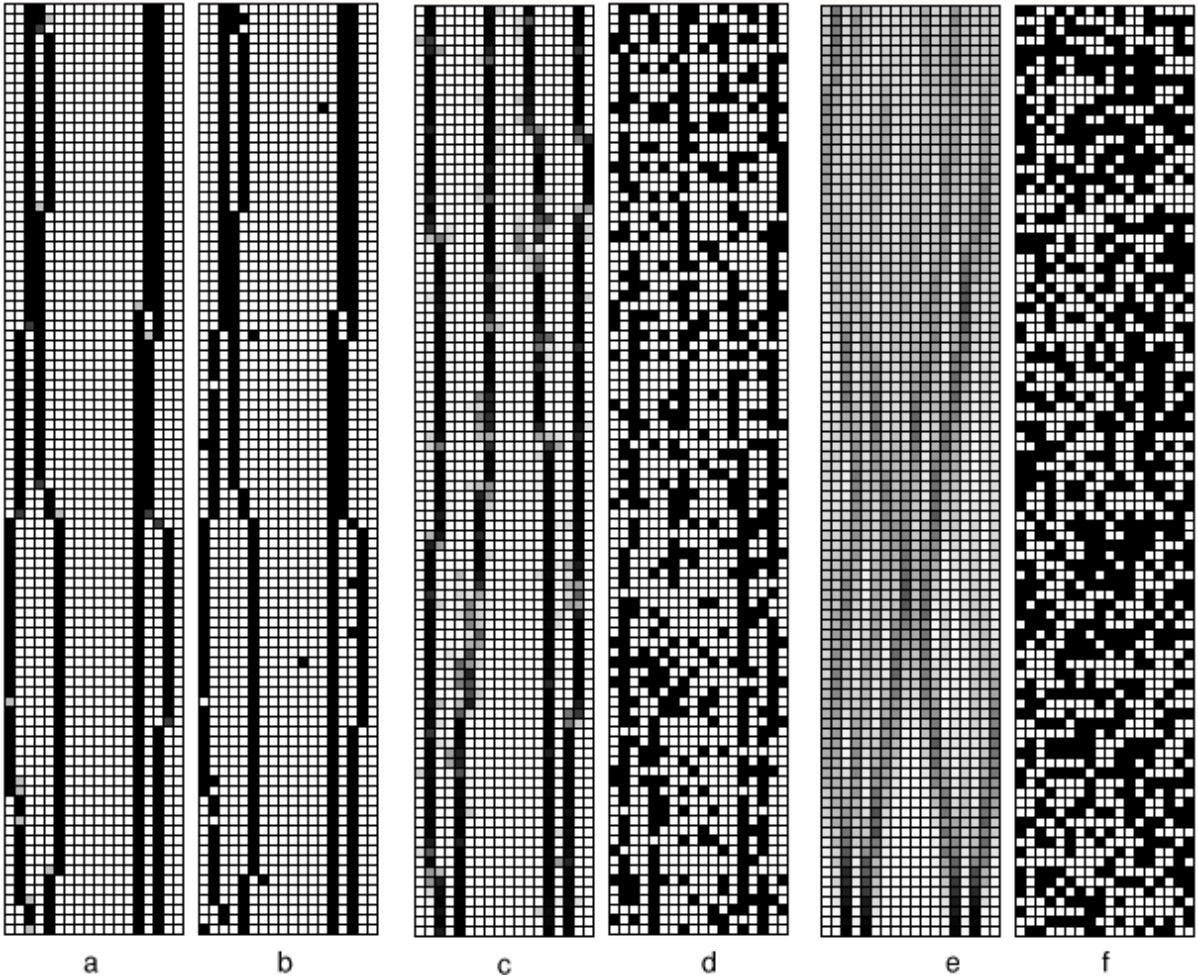}}
\caption{Pairs (a,b), (c,d) and (e,f), of (SI,
FI) plots for three different runs of the simulation.
 The initial state is a field eigenstate
with four particles and $\theta=4\pi/9$ in all three pairs. The pairs
have different values of $X$: $X=0.1$ for (a,b), $0.5$ for (c,d) and
$0.93$ for (e,f). }
\label{fig:4part}
\end{figure}

The same trend is apparent in figure \ref{fig:4part} which 
shows three runs with initial state a field  
eigenstate with 4 particles and with $\theta=4 \pi/9$.
The plots are again grouped in (SI, FI) pairs 
(a,b), (c,d) and (e,f) with  three different choices of
$X$, $X=0.1$, $0.5$ and $0.93$ respectively. 
The particles trajectories are close to being vertical 
 because $\theta$ is close to $\pi/2$
illustrating how $\theta$ controls the speed of the 
particles on the lattice. 

\section{Collapse of superpositions}

In order to fully justify the title ``dynamical collapse model'' 
it needs to be
shown that when only small numbers of degrees of freedom are involved,
the model behaves virtually indistinguishably from ordinary quantum
mechanics but that superpositions of ``macroscopically different''
states collapse rapidly.  We will present analytic and numerical
evidence that this is so -- to the extent that we can interpret
``macroscopic'' -- if we tune the parameter $X$ to be sufficiently
close to 1.

We can see that as $X$ gets close to 1, the realised field
configurations become more and more noisy until it is impossible to
discern any structure by eye (as seen for example in figure 
\ref{fig:2part} (f)).  Indeed when $X = 1$, the behaviour of the 
field is just
that of independent choices of 0 or 1 on each link with probabilities
$1/2$ and $1/2$. The state, on the other hand, decouples from the
field and evolves deterministically according to the standard unitary
evolution.  It seems likely that tuning $X$ to be close to 1 will
allow us to make the dynamics of the state as close as we like to
ordinary Schr\"odinger evolution for as long as we choose.  Since we
are aiming to make a comparison to ordinary quantum mechanics we study
only the SI in this section.

Let $X = 1 - \epsilon$ with $0 < \epsilon << 1$ and let  $\theta =
 \pi/2$ so the R-matrix is the identity and there is no ``Hamiltonian''
 evolution. Consider the simple case of a single link on which the
 state is the superposition $ |1\rangle + |0\rangle$ (suitably
 normalised). After $k$ time steps, the state will be $ X^{m_0 -m_1}
 |1\rangle + |0\rangle$ (suitably normalised) where $m_i$ is the
 number of times the field value $i$ is realised 
and $m_1 + m_0 = k$. This is the discrete 
 `quantum gambler's ruin' model considered in
\cite{Nakano:1994}. If the probabilities of 1
 and 0 were $1/2$ each then this would be governed by the simplest
 random walk on the integers ($m_0$ is the number of steps to the left
 and $m_1$ is the number of steps to the right).
 The expected value of the distance from the origin is of order the
 square root of the number of steps and so after $k$ steps the state
 is expected to be the superposition of the two original states with
 one suppressed (with respect to the other) by a factor $X^{\sqrt{k}}
 \approx \exp({-\sqrt{k}\epsilon})$.  When $k$ is large this means
 that one of the terms in the superposition is exponentially
 suppressed within a number of time steps of order
 $\epsilon^{-2}$. With this simple random walk, the superposition
 would always come back with full force because the walker always
 returns to the origin.  However the probabilities in the collapse
 model are not constant but depend on the state and the general theory
 of collapse models as stochastic processes in Hilbert space
 \cite{Ghirardi:1990, Nakano:1994} shows that in the limit of late times, the
 superposition is eventually suppressed.

For fixed $\epsilon$, we expect the decay time for a superposition of
two $\alpha$-eigenvectors to depend on the number of alpha values that
differ in the two configurations. For example, if the superposition is
of a state with $l$ particles on the left half of the lattice and a
state with $l$ particles on the right half, the decay time will scale
like $l^{-2}$.

\begin{figure}
\begin{minipage}[b]{0.5\linewidth} 
\centering
\includegraphics[width=6cm]{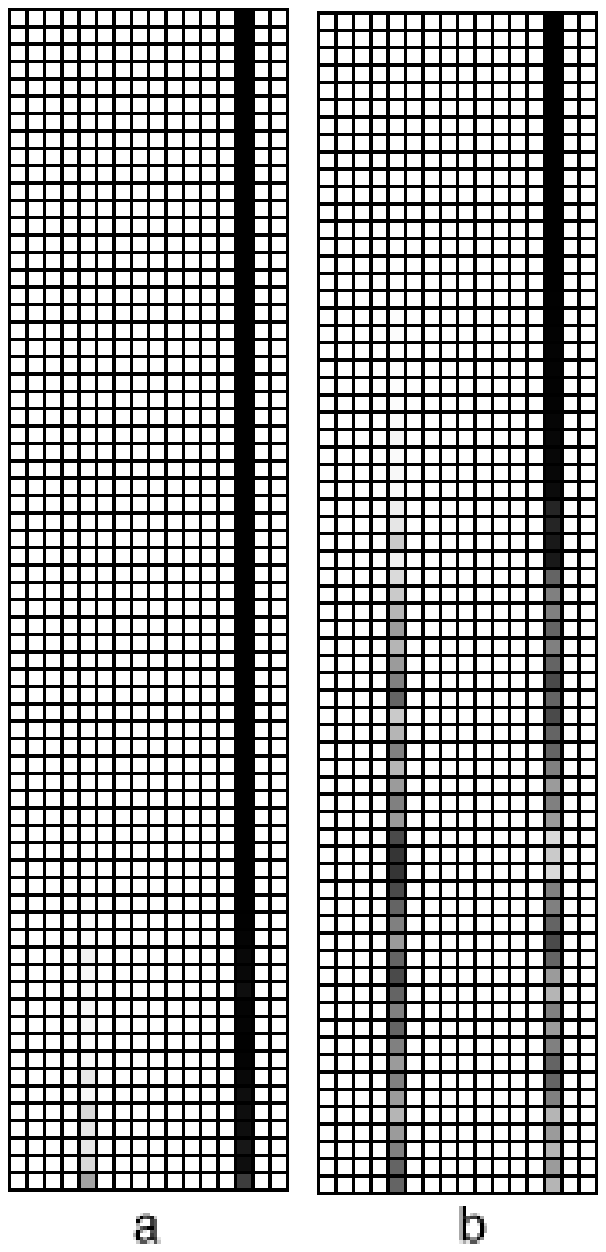}
\caption{Decaying 1 particle superpositions 
for different values of $\epsilon$: $\epsilon=0.25$ 
in (a) and $0.20$ in (b).}
\label{fig:1part_sup}
\end{minipage}
\hspace{0.5cm} 
\begin{minipage}[b]{0.5\linewidth}
\centering
\includegraphics[width=4cm]{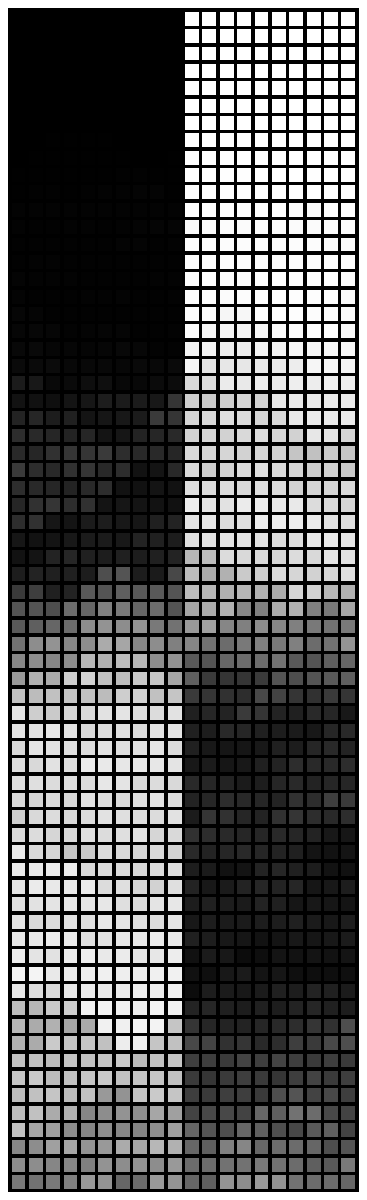}
\caption{Decaying 10 particle superposition with $\epsilon=0.05$. 
}
\label{fig:10part_sup}
\end{minipage}
\end{figure}

These expectations are supported qualitatively in figures 
\ref{fig:1part_sup} and \ref{fig:10part_sup}.
Figure \ref{fig:1part_sup} is two SI plots showing decaying single particle
superpositions
for two different values of $\epsilon$ ($\epsilon=0.25$, $0.20$) 
on an 8 vertex lattice.
Figure \ref{fig:10part_sup} is an SI plot of a run with a 10 particle
superposition initial state and $\epsilon=0.05$ on a 10 vertex lattice.  
$\theta =\pi/2$ in both figures so there is no Hamiltonian evolution.
We see that decreasing $\epsilon$ increases the decay time between 
figure \ref{fig:1part_sup} (a) and (b) significantly. The even 
smaller $\epsilon$ of $0.05$ would result in a 
much longer decay time for a 1 particle superposition but for 
a 10 particle superposition figure \ref{fig:10part_sup}
it results in a comparable decay time to figure  \ref{fig:1part_sup} (b). 

\begin{figure}[thb]
\epsfxsize=14cm \centerline{\epsfbox{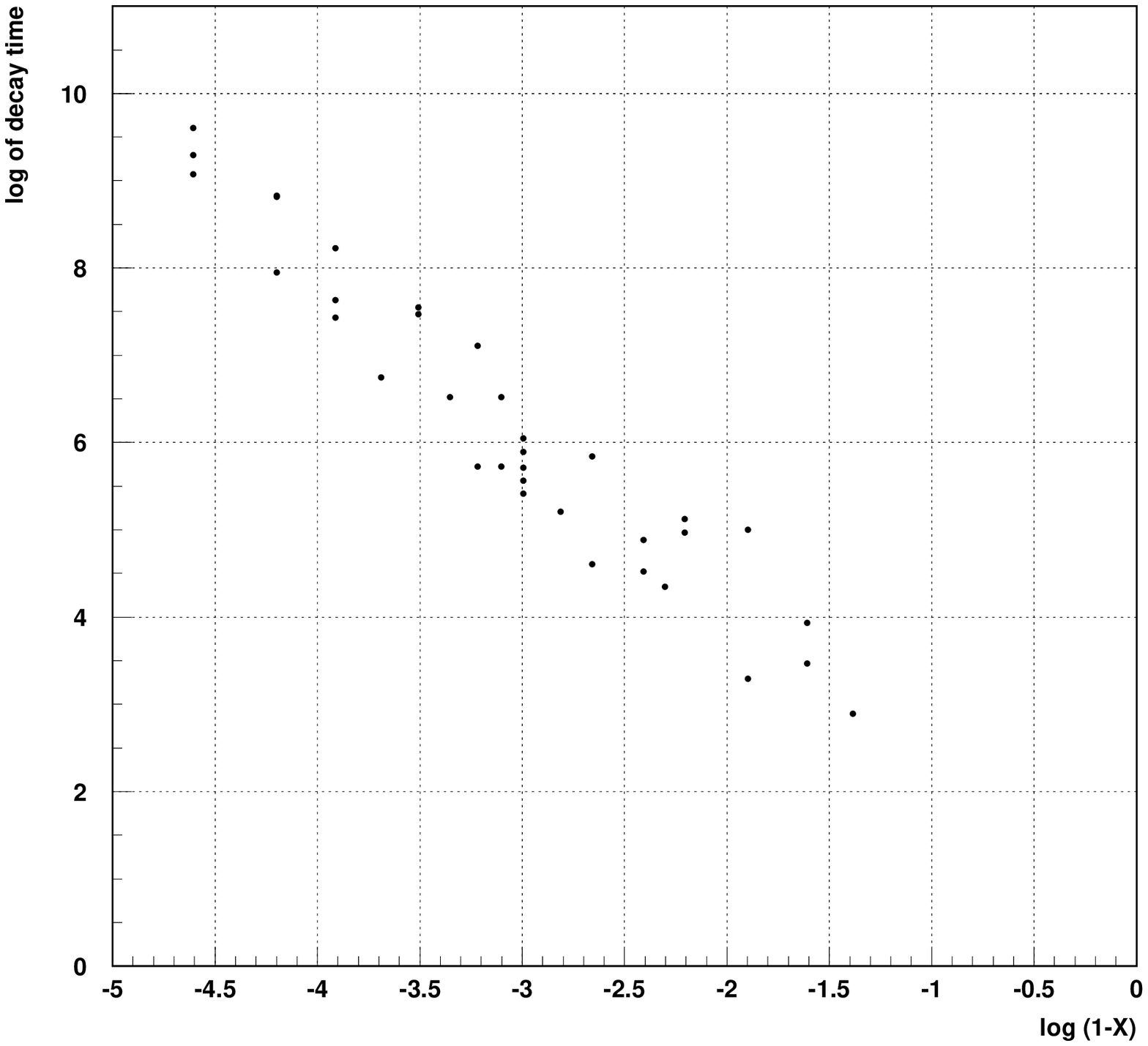}}
\caption{Plot of $\log(T_{decay})$ vs $\log(1-X)$. 
$T_{decay}$ is the time at which the squared modulus of the amplitude 
of one of the superposed states has
reached a value of $0.005$}
\label{plot1:fig}
\end{figure}

More quantitatively, from the analysis above, 
we expect that the ``decay time'', $T_{decay}$ scales like
$\epsilon^{-2}$ and figure \ref{plot1:fig} is a plot of over 35 runs
on an 8 vertex lattice with different $X$ values which supports this.
For each run, the superpositions were of 1 particle states, with 
equal initial amplitudes. The decay time is defined to be the time on the 
lattice at which the squared modulus of the
amplitude of one term of the superposition 
falls to one hundredth of its initial size (and the term 
remains suppressed after that, which is checked by running the simulation 
for much longer, not shown in the plots).

\begin{figure}[thb]
\epsfxsize=14cm \centerline{\epsfbox{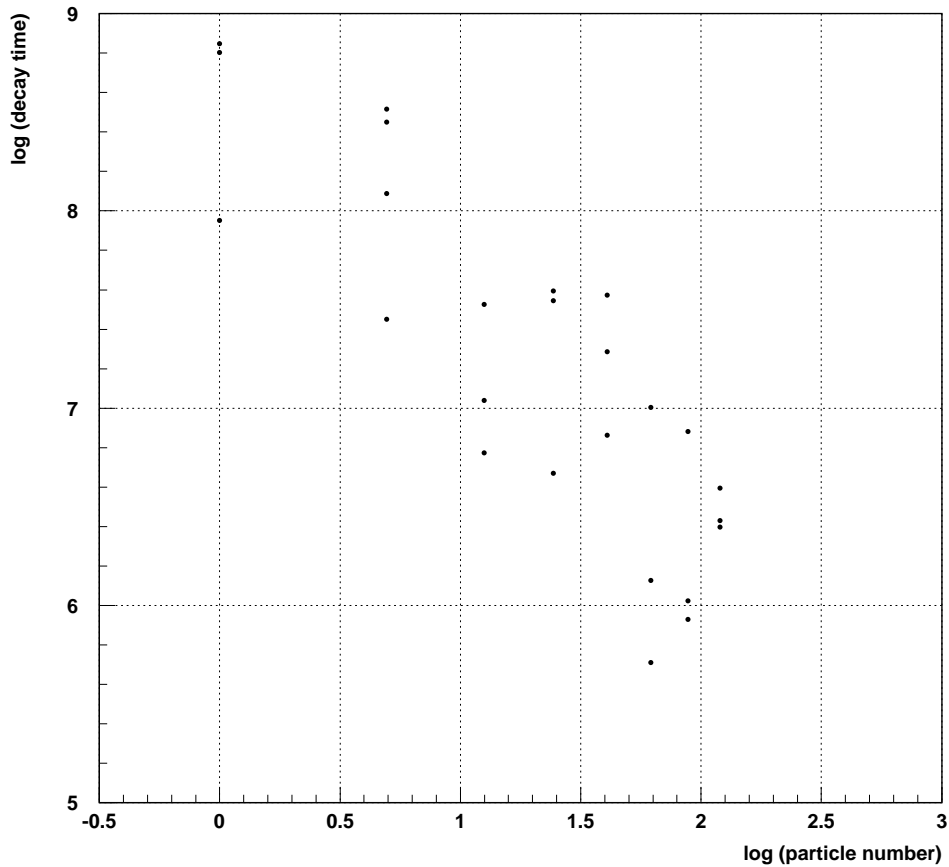}}
\caption{Plot of $\log(T_decay)$ vs $\log({\text{particle number}})$.}
\label{plot2:fig}
\end{figure}

 Figure \ref{plot2:fig} is a plot which  is broadly
consistent with our expectation for the dependence of the 
decay time on the particle number. The log of the decay time (as defined 
for figure \ref{plot1:fig}) of a superposition of two field eigenstates,
one with $l$ particles  on the left half of the lattice and one with 
$l$ particles on the right, is plotted against $\log{l}$
for a fixed value of $\epsilon = 0.015$. 
The decay times for the runs with small particle number, 
particularly for single particle superpositions, seem to 
be too  low for the fit. This is likely to  be because 
the value of $\epsilon$  is not small enough to be able to neglect
the ${\cal{O}}(\epsilon^2)$ terms.

The decay times for models with general R-matrix will depend on the
extent to which the dynamics and the collapse effects compete.  We can
see that if the R-matrix is very different from a permutation of basis
vectors (in our case a permutation corresponds to $\theta = 0$ or
$\pi/2$) and the jump operators are very close to ${\bf {1}}/\sqrt{2}$, 
the Hamiltonian dynamics will win the competition: the superpositions
that the jump operators are patiently suppressing with their tiny hits
would keep being reintroduced with a vengeance by each R-matrix
application. On the other hand an $\epsilon$ close to 1 will not allow 
any superpositions to persist, whatever the R-matrix is, 
illustrated by figure \ref{fig:2part}. 
Thus, for any R-matrix, an $\epsilon$ can be chosen that gives 
collapse of superpositions of field eigenstates but if 
$\epsilon$ is too small, superpositions will always reappear. 
An interesting question is whether, for a 
fixed R-matrix,  there is a critical value 
of $\epsilon$ between these two different sorts of behaviour and 
what the theory looks like there. 

Another interesting  direction is to consider the continuum limit 
in which the lattice spacing $a \rightarrow 0$ with $L\equiv Na$ 
held fixed and 
simultaneously $\epsilon \rightarrow 0$. 
Looking at the ``bare continuum limit'' taken 
by Destri and deVega, suggests that $\epsilon$ 
behaves as $\epsilon = {\cal{O}}(\sqrt{a})$. 
In that case, $\theta = {\cal{O}}(\epsilon^2)$ and for 
small $\epsilon$ the R-matrix will indeed be very close to a permutation. 

\section{Coarse graining and renormalisation}

At first sight it appears that the requirement that superpositions of
single particles persist for a long time means that the FI is
untenable.  Since the decay time is inversely
proportional to $\epsilon^2$, $\epsilon$ must be tuned to be very
small and the probabilities of 1 and 0 on each link are both $1/2 +
{\cal{O}}(\epsilon)$. This means that the fundamental field configurations
become extremely noisy. We suggest that by a process of coarse
graining and renormalisation of the field values, the FI can remain
viable and the fields display interesting structure.

For $\epsilon =0$, coarse graining give us the statistics of
discretised Gaussian white noise, {\it i.e.}  the binomial
distribution: if we coarse grain the field values over square blocks
of the lattice containing $M = m \times m$ lattice vertices -- that is
$2m^2$ links -- the mean of the distribution over the averaged block
field values is $\mu = 1/2$ and the variance is $\sigma^2 = 1/(8M)$.
This is independent of the state.

For $\epsilon > 0$ and the vacuum state $|00\dots 0\rangle$, with
particle number preserving R-matrix, the state is preserved and the
stochastic field dynamics is an independent choice on each link of 0
with probability $1/(1 + X^2)$ and 1 with probability $X^2/(1 + X^2)$.
For the coarse grained field we again have the statistics of a
binomial distribution with mean $\mu = X^2/(1+ X^2) = 1/2 +
{\cal{O}}(\epsilon)$ 
and variance $\sigma^2 = X^2/(2 (1+ X^2)^2 M)$. Let us call
this the vacuum distribution for a particular $\epsilon$. We want to
``subtract'' this fluctuating vacuum background and see if the
remaining field has interesting features. The difference, for each
block, between the mean of the vacuum distribution on coarse grained
field values and the mean of the distribution corresponding to a
general initial state will be of order $\epsilon$ and so to ``see''
these differences we need to renormalise the field.

Given a block size $m$ we define a renormalised field in the following
way. Let $\alpha_m$ be the average value of the field in an $m \times
m$ block. Then the renormalised field in that block is
\be \alpha_m^R = \epsilon^{-1}(\alpha_m - X^2/(1+X^2)) \; .  \ee
Although the underlying (bare) field values are only 0's and 1's and
the averaged field values therefore lie between 0 and 1, the
renormalised values extend over a larger and larger range 
as $m$ increases (with large field values
exponentially unlikely). 

We want this renormalised field not to be swamped by vacuum
fluctuations and so we need the square root of the variance of the
distribution to be much smaller than the difference in the means, which
implies that $m \epsilon \gg 1$. Note that if a continuum limit 
were taken with $a$ the lattice spacing going to zero, it would be 
reasonable for $m a$ -- the physical ``discrimination scale'' -- 
to remain fixed. In which case 
$m = {\cal O}(a^{-1}) = {\cal O}(\epsilon^{-2})$.

We see that is hard to simulate this coarse grained field dynamics for small
$\epsilon$ because our lattice size doesn't allow the large block size that
would be needed. Instead we offer the following argument as to why it
is plausible that the theory is non-trivial.

As a substitute for showing the field values themselves, we consider
renormalised and enhanced probabilities on the links. Consider, at
some stage in the dynamics, the link $l$ which is one of the outgoing
links from the vertex that has just been evolved over. Then the state
on the surface through $l$ is given by 
\eqref{current.eq} and the probability that the field
will be 1 on $l$ is given by \eqref{probone.eq}. 
When $\epsilon$ is very small, this probability 
is within order $\epsilon$ of the
vacuum value $X^2/(1 + X^2)$. Subtracting off this vacuum probability,
and enhancing the contrast by dividing this by $\epsilon$, we obtain
the value $|b|^2$ (to order $\epsilon$).  This is exactly the quantity
we show in the SI plots (it is the stuff).

We are led to the conclusion that when the state shows structure on 
some coarse grained scale, it is likely that the
renormalised coarse grained field also will and we find a strong
connection between the Field and State Interpretations.  

\section{Discussion}

We have presented analysis and numerical simulations of lattice models
that can be thought of as dynamical collapse models on a discrete
spacetime with a fixed causal structure.  At the very least the
discreteness and finiteness of the model allows us to see clearly the
elegant mechanism of the dynamical localisation models of Ghirardi,
Rimini, Weber and Pearle at work. The competition between collapse
fuelled by the stochasticity and the Hamiltonian evolution is seen
clearly: the jump operators suppress superpositions in the field
eigenstates and the R-matrices reintroduce them at each step.

We have highlighted the choice between the State Interpretation and
the Field Interpretation. We have emphasised the fact that the 
SI struggles to be covariant in the presence of a non-trivial 
causal structure due to the dependence of local
quantities on a choice of hypersurface through the local region 
on which the quantity is defined. The FI on the other hand is 
manifestly covariant and respects the causal structure. 

We have presented evidence that the Field Interpretation can remain
interesting, through the noise, by coarse graining and renormalising
the fundamental field configurations. This clearly has some attractive
features: there really {\it are} vacuum fluctuations at some
fundamental scale but coarse graining washes them out. It remains to
be discovered whether, for a value of $\epsilon$ close enough to zero
to maintain microscopic superpositions, the renormalised coarse
grained field has interesting structure. The fact that the linear size
of the coarse graining blocks must grow as $\epsilon$ gets small means that
we cannot look for this directly in our simulations on small
lattices. However, we have shown that this question is closely related to the
behaviour of the state on coarse grained scales: if the SI predicts
interesting coarse grained structure then so does the FI.

An interesting issue that arises for collapse models especially in the
relativistic context, is that of energy conservation. The consensus seems to 
be that collapse models -- and non-unitary evolution in general -- 
if local and causal will tend
to violate energy-momentum conservation, more or less severely. 
In the GRW model, that can be seen simply by noting that the collapses 
tend to cause wavepackets to narrow in position and so to
broaden in  momentum. These conclusions, however, have been 
drawn in the SI in which the state is used to calculate 
the various quantities. The question of whether energy-momentum 
conservation holds, and if not how large the violations will be,
will have to be readdessed in the FI where it is likely to 
take on rather a different 
character. For example, the fundamental field configurations fluctuate 
wildly (in the continuum, they would be non-differentiable and perhaps 
not even continuous) and so a microscopic definition of energy, say, 
would not be possible. A definition depending on some coarse graining 
would be necessary. We leave these questions open for now. 

The FI suffers from the defect that it is necessary to know the state
on a spacelike surface in order to predict the future behaviour of the
field. It seems strange therefore not to ascribe reality to the state as
well. Indeed, the interpretation in which {\it both} the stochastic
quantity (here the field) and the state vector are real 
has been proposed by Di\'osi in many works over the 
years (already implicitly in \cite{Diosi:1988a}). 
There is no getting around the reality of the state if one wants 
to maintain a Markovian dynamics, but if one is prepared to
give up the Markovian condition, it might be possible to show that the
state on a spatial surface is actually determined, for all practical
purposes, by the realised field values that have occured in the past
back to some time depth $T_{memory}$. We will report on work bearing
on this issue in a future paper.

In the original GRW model there are two parameters, essentially the
 microscopic decay time, $T_{decay}$, for collapse of a superposition
 of a single particle in two different positions and the spatial
 ``discrimination'' scale, $X_{discrim}$, which is the smallest
 distance that the collapse can resolve. In the GRW model the values
 $T_{decay} = 10^{16}s$ and $X_{discrim} = 10^{-5} cm$ are chosen.
 These two parameters appear in the current setting incarnated as
 the lattice spacing and the parameter $\epsilon$.  Given a fixed
 lattice spacing, demanding a particular value for $T_{decay}$ fixes
 the discrimination scale in the following way.

Let $T_0$ be the
 lattice spacing in time so that $X_0 = c T_0$ is the lattice spacing
 in space. Since the decay time scales like $\epsilon^{-2}$ we have
 that $T_{decay} = k T_0 \epsilon^{-2}$ where k is a dimensionless 
number that depends on the dynamics {\it i.e} on the R-matrix.  
The discrimination length scale is 
$X_{discrim} = m X_0$ where $m$ is the number of vertices in one 
coarse graining block dimension. We know that $K \equiv m \epsilon \gg 1$.
Putting  these together we obtain 
\begin{eqnarray}
X_{discrim} &= K k^{-\frac{1}{2}} X_0 \left({\frac{T_{decay}}{T_0}}
\right)^{\frac{1}{2}}\\
            &= Kk^{-\frac{1}{2}} \times 10^{-3} {\text cm} 
\end{eqnarray}
if we choose $T_{decay} \sim 10^{16} s$ as in the GRW and CSL models
and make a ``natural'' choice for the
lattice spacing, namely the Planck scale ($X_0 = 10^{-33} cm$). 
$X_{discrim}$ is in danger of being too large, because $K \gg 1$.
But if the R-matrix can be tuned to compete with the collapse 
mechanism to the extent that $k$ is large enough so that 
$K/\sqrt{k} \sim 10^{-2}$ then we can obtain the
 GRW value. There is therefore a hint that one of the
extra physical constants of the GRW/CSL models can be eliminated by
invoking a fundamental spacetime discreteness at the Planck scale.

This is to be regarded as merely a hint for several reasons including
the fact that the diamond lattice used here is not Lorentz invariant.
In order to produce a model that has a chance of being Lorentz
invariant we would either have to take the continuum limit of the
lattice models (and deal with the issues that arise there:
non-physicality of the ``bare'' particles, the fact that the physical
vacuum contains infinite numbers of bare particles {\it etc.}) or
build a collapse model on a Lorentz invariant discrete structure to
which Minkowski spacetime is an approximation.  The only known example
of the latter is a causal set, which can be thought of as a discrete
``random sampling'' of the causal structure of Minkowski spacetime
\cite{Bombelli:1987aa, Dowker:2003hb}.  While it seems possible
formally to write down a collapse model on a general causal set, with
Hilbert spaces on the links and ``local'' evolution and collapse
rules, the analysis and simulation of such a model is immensely more
challenging. General models of this type have been considered
\cite{Blute:2001zd} in which the jumps are considered to be
interventions of some external agent on an open quantum system. In
\cite{Hawkins:2003vc} the dual situation where Hilbert spaces live on
the vertices is considered from a density matrix perspective.

Speculating much further, one motivation for studying collapse models
is to find an observer independent approach to quantum theory that can
be applied to the problem of quantum gravity, and here it is harder to
see how this might work. In the causal set approach to quantum
gravity, the major outstanding problem is to find a quantum dynamics
that will produce causal sets that are well-approximated by continuum
spacetimes. The models investigated above could be viewed as
stochastic models for generating causal sets from the diamond lattice
in the following way: when the field value is 0 delete the link from
the lattice and when it is 1 keep the link. This is therefore a
specific, causal example of the type of models described in
\cite{Markopoulou:1997wi}.

One difficulty with this proposal for dynamically generating causal
sets is that the only causal sets that can arise in this way (by
deleting links from the diamond lattice) are very special and look
nothing like the causal sets which have Minkowski space (or any other
continuum spacetime) as an approximation. For example, each element
can have at most two future and two past links.  In order to give
ourselves a fighting chance of generating in this way a causal set
that looks like a continuum spacetime, we would need, rather, to
consider deleting {\it relations} from a causal set (a relation is
anything implied by the links -- including the links themselves -- by
transitivity). It suffices, then, to consider deleting relations from
the causal set that contains all possible relations: the totally
ordered causal set known as the chain.

It isn't hard to to see how one might form models along these lines:
for each positive integer $N$, consider a collapse model for a
$\{0,1\}$ valued field living on the $N(N-1)/2$ relations of the $N$
element chain.  This generates a probability distribution on all
labelled $N$ element causal sets.  We impose the condition that all
these distributions must be consistent with each other: the induced
probability distribution on $N$ element causal sets from the
distribution for $N+1$ must agree with the previous one.  We also
require that the distributions must be generally covariant: the
probability of two finite labelled causal sets which are isomorphic
must be equal.  The models will be thus be examples of the classical
sequential growth models considered in \cite{Rideout:1999ub} (indeed
our $\{0,1\}$ field is reminiscent of the ``Ising matter''
interpretion given therein) but without the condition of Bell
Causality, which we do not want to impose in quantum gravity because
it would lead to the Bell Inequalities. One problem with these models
without the Bell Causality condition is that there will be a huge
number of them. Without the discovery of an 
additional well-motivated physical
criterion to limit the possibilities, there seems little reason to
expect they have anything to do with quantum gravity.

\section{Acknowledgments}

We would like to thank Lajos Di\'osi, Joe Henson, Gerard Milburn, Trevor
Samols and Rob
Spekkens for helpful discussions. We also thank Paul Dixon for help
with the simulations.

\bibliography{../../../Bibliography/refs} \bibliographystyle{JHEP}

\end{document}